\documentclass[aps,prl,twocolumn,showpacs,superscriptaddress,groupedaddress]{revtex4-1}
\usepackage{dsfont,amsthm,amsbsy}
\usepackage{amssymb}
\usepackage{amsmath}
\usepackage{bbm}
\usepackage{graphicx}
\usepackage{epstopdf}
\usepackage{subfigure}
\usepackage{natbib}
\usepackage{epsfig}
\usepackage{amsfonts}
\usepackage{mathrsfs}
\usepackage{sidecap}
\usepackage{lipsum}
\usepackage[toc,page,title,titletoc,header]{appendix}
\usepackage[colorlinks,linkcolor=blue,citecolor=blue,anchorcolor=blue, urlcolor=blue]{hyperref}
\usepackage{hyperref}
\usepackage{dsfont,amsthm,amsbsy}
\usepackage{resizegather}
\usepackage{tikz}
\usepackage{float}
\usepackage{mathbbol}
\newcommand{\be}{\begin{equation}}
\newcommand{\ee}{\end{equation}}
\newcommand{\bea}{\begin{eqnarray}}
\newcommand{\eea}{\end{eqnarray}}
\newcommand{\bi}{\begin{itemize}}
\newcommand{\ei}{\end{itemize}}
\newcommand{\bc}{\begin{center}}
\newcommand{\ec}{\end{center}}

\hypertarget{letter}{}
\begin{document}

\title{Movable but not removable band degeneracies in a symmorphic crystal}

\author{Mariana Malard$^{1,2}$, Paulo Eduardo de Brito$^{1}$, Stellan \"Ostlund$^{2}$, and Henrik Johannesson$^{2,3}$}
\affiliation{$\mbox{}^1$University of Brasilia, 70904-910, Brasilia-DF, Brazil}
\affiliation{$\mbox{}^2$Department of Physics, University of Gothenburg, SE 412 96 Gothenburg, Sweden}
\affiliation{$\mbox{}^3$ Beijing Computational Science Research Center, Beijing 100094, China}

\begin{abstract}
Crossings of energy bands in solids that are not pinned at symmetry points in the Brillouin zone and yet cannot be removed by perturbations are thought to be conditioned on the presence of a nonsymmorphic symmetry. In this Letter we show that such band crossings can actually appear also in a symmorphic crystal.  A study of a class of tight-binding multiband one-dimensional lattice models of spinful electrons reveals that chiral, time-reversal and site-mirror symmetries are sufficient to produce such \emph{movable but not removable} band degeneracies.
\end{abstract}


\maketitle

\emph{Introduction} - Level crossings $-$ the appearance of degeneracies in the spectrum of a Hamiltonian $-$ underlie a variety of phenomena, from quantum phase transitions \cite{Sachdev2001} to properties of topological semimetals \cite{Armitage2018}. The non-crossing rule by von Neumann and Wigner \cite{Neumann1929} here gets circumvented by the presence of one or several symmetries which inhibit level repulsion. When a level crossing occurs through tuning a control parameter, the resulting degeneracy is said to be \emph{accidental}; else, if symmetry alone dictates the presence of the degeneracy, it is commonly called \emph{symmetry-enforced}.

Level crossings, or ``nodes", play a particularly important role in the theory of electronic band structures of solids \cite{Martin2004}. Whereas the possibility of accidental band degeneracies was pointed out early on \cite{Herring1937}, only rather recently have their physical implications been more systematically investigated, leading to the discovery of Weyl semimetals  \cite{Murakami2007,Wan2011}. Symmetry-enforced degeneracies, on the other hand, have long played a key role in band theory. Typically pinned at high-symmetry momenta in the Brillouin zone (BZ) \cite{Bouckaert1936}, they form the ``essential" degeneracies well known from text books  \cite{DresselhausBook}. A seemingly unique situation occurs in the presence of a nonsymmorphic symmetry, i.e. when the crystal is invariant under a point group transformation combined with a {\em nonprimitive} lattice translation \cite{DresselhausBook}. In this case, the electronic bands form a connected net \cite {MichelZak1999} and while the resulting nodes cannot be lifted by symmetry-preserving perturbations, their location can be moved in the BZ by the same perturbation, leading to the notion of \emph{movable but not removable degeneracies}.

The degeneracies which emerge from nonsymmorphic symmetries have come to play a crucial role in the theory of Dirac \cite{Young2012,Young2015} and nodal line \cite{Fang2015} semimetals. It has recently been shown that they may appear also in other unconventional band structures, leading to nodal chains \cite{Sigrist2016} and surface modes with ``hourglass" dispersions \cite{Wang2016}. The mobility of these nodes throughout the BZ $-$ when unconstrained by other symmetries $-$ suggests that their robustness against perturbations is linked to a global topological invariant \cite{Zak2002,Zhao2016}. This is different from the movable accidental nodes in Weyl semimetals which are endowed with only local topological protection \cite{Murakami2007}. For extended discussions of symmetry-enforced nodal phenomena in semimetals, and also in unconventional superconductors, see Refs. \onlinecite{A,B}.

Given the importance of symmetry-enforced and yet unpinned degeneracies, one may inquire whether similar level crossings can appear also in a symmorphic crystal, characterized by invariance under point group transformations and {\em primitive} lattice translations \cite{DresselhausBook}. In this Letter we show that this is indeed possible. Specifically, we show that a pair of movable but not removable nodes exists in the multiband spectra of a class of symmorphic tight-binding chains of spinful electrons possessing chiral, time-reversal and site-mirror symmetries. When perturbed, these nodes move symmetrically in the BZ, conspicuously making them akin to Weyl nodes \cite{Murakami2007}, with the crucial difference that here they cannot be pairwise annihilated through a perturbation which respects the underlying symmetries. Relevant for applications, realizations of the investigated class of models may be engineered from a quantum wire supporting spin-orbit interactions of arbitrary strength. The fact that the symmorphic mirror symmetry enforces movable but not removable nodes already in one spatial dimension allows for a simple and transparent analysis. We shall build our argument starting from a chain of spinless fermions, and then show how our result emerges by by bringing in spin.

\emph{Spinless chains with chiral, time-reversal and inversion symmetries} - Consider a translational invariant one-dimensional (1D) lattice with $r\in2\mathds{N}$ sites per unit cell, distributed between two sublattices, one formed out of the odd-labelled sites and the other from the even-labelled sites. The chain is populated by spinless fermions with nearest-neighbor hopping only. The $r \times r$ Bloch matrix describing the system in the spinor representation introduced in the Supplemental Material (SM) \cite{Malard2017} has the general form
%%%%%%%%%%%%%%%%%%%%%%%%%%%%%%%%%%%%%%%%%%%%
\begin{equation} \label{Hk}
{\cal H}(k)\,=\, \begin{bmatrix}
    0 & Q(k) \\
    Q^{\dagger}(k) & 0 \\
\end{bmatrix},
\end{equation}
%%%%%%%%%%%%%%%%%%%%%%%%%%%%%%%%%%%%%%%%%%%%
where $Q(k)$ is the matrix containing the hopping amplitudes between the two sublattices. The model supports chiral symmetry, i.e. ${\cal S}\,{\cal H}(k)\,{\cal S}^{-1}\!=\!-{\cal H}(k)$, with ${\cal S}\!=\!\sigma_{z}\otimes\mathbb{1}_{r/2 \times r/2}$ the matrix implementing the chiral transformation. In addition, we impose time-reversal symmetry,  ${\cal T}{\cal H}(k){\cal T}^{-1}\!=\!{\cal H}^{\ast}(-k)$ with ${\cal T}\!=\!\mathbb{1}_{r \times r}$, implying real hopping amplitudes, which we take to be positive.

There are two ways in which a tight-binding chain with nearest-neighbor hopping may be invariant under inversion. They differ by the inversion point being located on the bond between two sites - ``bond-inversion" - or on a site - ``site-inversion". Fig. 1 illustrates both situations for $r=2,4$, with larger unit cells easily represented by repeating the underlying pattern. As seen in Fig. 1, a chain with two sites per unit cell supports only bond-inversion symmetry (chain $2b$), while for larger unit cells both types of symmetries are possible. Chain $2b$ corresponds to the well known spinless Su-Schrieffer-Heeger (SSH) model \cite{Asboth2016}.
%%%%%%%%%%%%%%%%%%%%%%%%%%%%%
%%%%%%%%%%%%%%%%%%%%     FIGURE 1
%%%%%%%%%%%%%%%%%%%%%%%%%%%%%%%%%%%%%%%%%%%%%%%%%%%%%%%%%%%%%%%%%%%%%%%%%%%%%%%%
\begin{figure}
\includegraphics[scale=0.5]{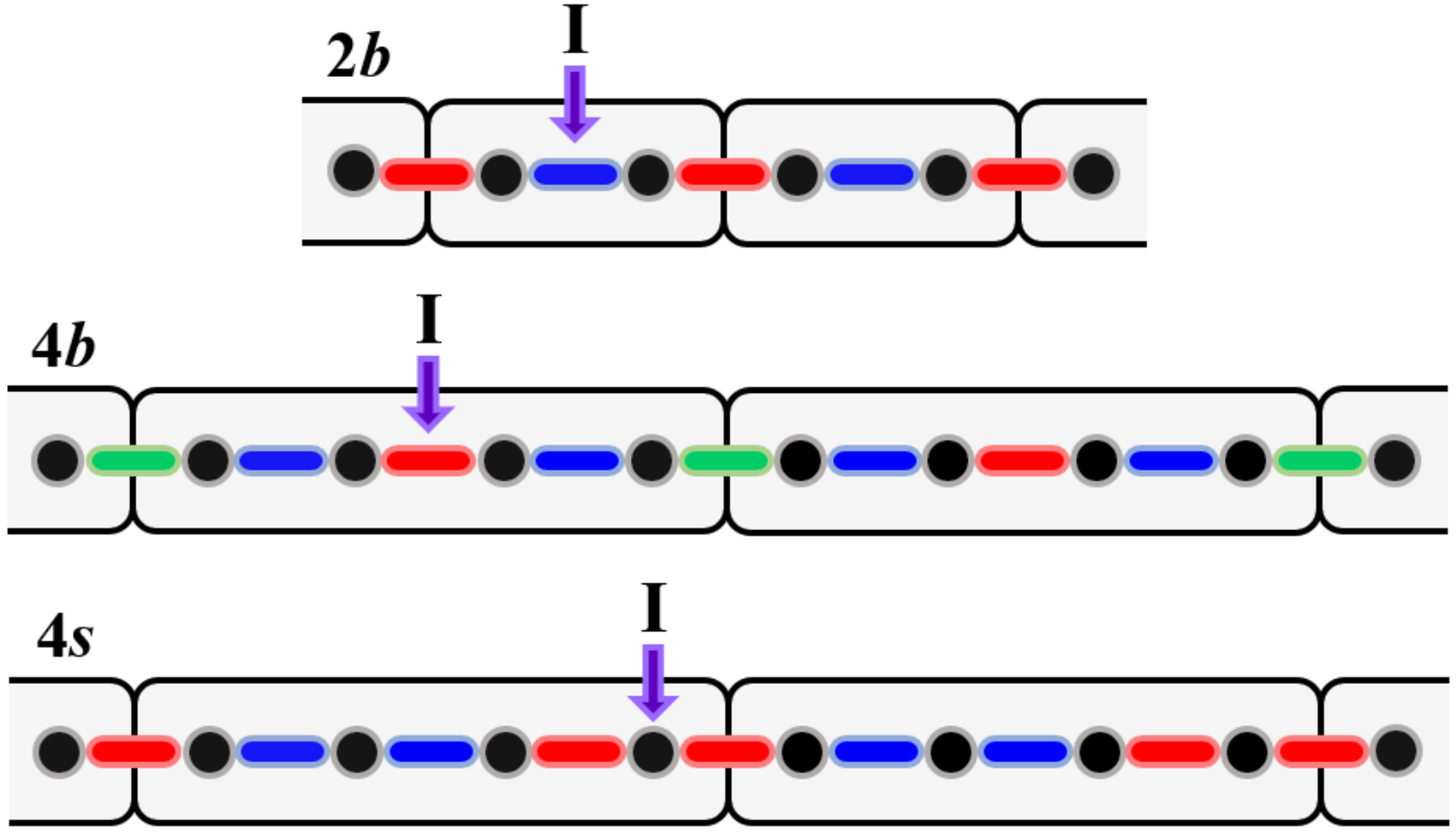}
\caption{(Color online) Chains with a bond-inversion point $-$ $2b$, $4b$ $-$ and with a site-inversion point $-$ $4s$. The colored segments represent bonds with different strengths. The inversion point is indicated by I.}
\end{figure}
%%%%%%%%%%%%%%%%%%%%%%%%%%%%%%%%

In the following we will analyze the cases with $r>2$. Our goal is to establish the conditions under which the gap closes through the appearance of a zero-energy degeneracy. Given that chiral symmetry forces the spectrum of ${\cal H}(k)$ to be symmetric around zero energy \cite{Asboth2016}, the existence of such a node is guaranteed if the spectrum has at least one zero eigenvalue. The latter requirement is fulfilled if $\text{det}[Q(k)]=0$. The $Q$-matrices for the chains in Fig. 1 with $r\!=\!4$ read
%%%%%%%%%%%%%%%%%%%%%%%%%%%%%%%%%%%%%%%%%%%%
\begin{equation} \label{Q4}
Q^{4b}\,=\, \begin{bmatrix}
    a & cz \\
    b & a \\
\end{bmatrix},\qquad\qquad Q^{4s}\,=\, \begin{bmatrix}
    a & bz \\
    a & b \\
\end{bmatrix},
\end{equation}
%%%%%%%%%%%%%%%%%%%%%%%%%%%%%%%%%%%%%%%%%%%%
where $z=e^{-ik}$ with $k \in [-\pi, \pi]$, and $a$, $b$, $c$ are the hopping amplitudes along the blue, red, and green bonds, respectively \cite{Malard2017}. The condition $\text{det}[Q(k)]=0$, subject to $|z|=1$, implies in each case: $z=z^{4b}=a^{2}/(bc)$ if $a^{2}=bc$; $z=z^{4s}=1$ for any $a$ and $b$. Since $z=e^{-ik}$, in both cases the node is located at $k=0$, a consequence of $a$, $b$ and $c$ being real numbers. The crucial difference comes from the constraint imposed on the hopping amplitudes, in the case of bond inversion, or lack thereof, in the case of site inversion: Bond-inversion symmetry, when combined with chiral and time-reversal symmetries, leads to an accidental node, while with site-inversion symmetry the degeneracy becomes unavoidable. This conclusion immediately generalizes to a unit cell with $r\!>\!4$ sites.

To prove that the combination of chiral ($S$), time-reversal ($T$) and site-inversion ($I$) symmetries enforces a $k=0$ node, we consider the site-inversion transformation $I(k)\!=\!{\cal I}(k)\times\!\parallel$, where the ``hard wall" operator $\parallel$ reverses momentum and ${\cal I}(k)$ is an $r\! \times \!r$ matrix acting on the intracell positions,
%%%%%%%%%%%%%%%%%%%%%%%%%%%%%%%%%%%%%%%%%%%%
\begin{equation} \label{Ik}
{\cal I}(k)\,=\, \begin{bmatrix}
    R_{1}(k) & 0 \\
    0 & R_{2}(k) \\
\end{bmatrix}.
\end{equation}
%%%%%%%%%%%%%%%%%%%%%%%%%%%%%%%%%%%%%%%%%%%
The forms of the $R_{1}(k)$ and $R_{2}(k)$ matrices depend on the size of the unit cell. If $I(k)$ is a symmetry transformation, then ${\cal H}(k)$ must satisfy ${\cal I}(k)\,{\cal H}(-k)\,{\cal I}^{-1}(k)\!=\!{\cal H}(k)$ \cite{Malard2017}. It follows, using Eqs. \!(\ref{Hk}),(\ref{Ik}), and the identity ${\cal I}^{-1}(k)\!=\!{\cal I}(-k)$, that $R_{1}(k)\,Q(-k)\,R_{2}(-k)\,=\,Q(k)$. With $r=4$ sites per unit cell, $R_{1}(k)\!=\!z\,\text{adiag}(1\,1), R_{2}(k)\!=\!z\,\text{diag}(1\,z^{\ast})$ \cite{Malard2017}, with the symbol $\text{diag}$ ($\text{adiag}$) denoting a diagonal (anti-diagonal) matrix and, as before, $z=e^{-ik}$. Assuming a generic $Q(k)$ with $r=4$, it follows that the $k$-independent parameters appearing in $Q(k)$, call them $q_{ij}$, must satisfy $q_{21}\!=\!q_{11}$ and $q_{12}\!=\!q_{22}$. This confirms that $Q^{4s}$ in Eq. (\ref{Q4}) is the most general matrix describing a spinless $STI$-invariant chain with $r=4$ sites per unit cell. Again, the procedure applies to an arbitrarily large unit cell with $r>4$ once the corresponding $R_{1}(k)$ and $R_{2}(k)$ have been obtained \cite{Malard2017}.

One can now understand how the noncrosssing rule is bypassed in the spinless $STI$-invariant chain. In order to avoid level repulsion, states must carry distinct quantum numbers. This requirement is satisfied by $S$ which prescribes that degenerate zero-energy states are eigenstates of the chiral operator with opposite eigenvalues \cite{Asboth2016}.  Still, $S$-symmetry alone only paves the way for the appearance of an accidental degeneracy. Adding $I$-symmetry constrains the Bloch matrix in such a way that a nodal solution exists in the whole parameter space. By enforcing real hopping amplitudes, $T$-symmetry pins the node at $k=0$. As we shall see, adding spin creates a pair of Kramers related nodes with the striking effect of unpinning them, without disrupting the symmetry enforcement.

A final remark on the spinless case: At a first glance, the $k$-dependance of $I$ might appear as a signature of a nonsymmorphic transformation, in which case our inversion would actually be a glide operation \cite{DresselhausBook}. This is not the case: By the crystallographic definition, the $k$-dependence of a nonsymmorphic transformation is along the direction parallel to the mirror plane \cite{Konig1997}. In the case of a 1D system, $k$ is, by construction, perpendicular to the plane of inversion. The $k$-dependence of $I$ instead comes about from the lack of invariance of the unit cell under the site-inversion transformation which, in turn, stems from the offset between the inversion point and the center of the cell (see Fig. 1). This is a feature of site-inversion which does not occur with bond-inversion. Using the property ${\cal I}(-k)={\cal I}^{-1}(k)$, it can also be seen that $I^{2n}(k)=1\!\!1$, and thus $I^{2n+1}(k)=I(k)$, with $n=1,2,...$. This means that, unlike a nonsymmorphic transformation, $I(k)$ cannot be iterated to eventually produce a full translation $e^{ik}1\!\!1$. For discussions of other lattice models with symmorphic $k$-dependent symmetry transformations, see Refs. \onlinecite{Furusaki,Schnyder,Brzezicki2017}.

\emph{Spinful chain with chiral, time reversal and site-mirror symmetries} - Let us consider again the minimal $4s$-chain which, in the spinful case, can be represented as in Fig. 2. The matrix $Q^{4s}$ from Eq. (\ref{Q4}) is now replaced by
%%%%%%%%%%%%%%%%%%%%%%%%%%%%%%%%%%%%%%%%%%%%
\begin{equation}
\tilde{Q}^{4s}\,=\,\begin{bmatrix}\label{Q4sSpinful}
    A & B^{\ast}z \\
    A^{\ast} & B \\
\end{bmatrix},
\end{equation}
%%%%%%%%%%%%%%%%%%%%%%%%%%%%%%%%%%%%%%%%%%%%
where the hopping amplitudes $a$ and $b$ became $2\times2$ matrices $A$ and $B$ whose diagonal (off-diagonal) entries account for hoppings with equal (flipped) spin \cite{Malard2017}. An experimental realization of both the spin-conserving and spin-flipping terms in ${\tilde{Q}}^{4s}$ may be found in a quantum wire with spatially modulated Rashba and uniform Dresselhaus spin-orbit interactions \cite{Malard2016}. The Bloch matrix, given by Eq. (\ref{Hk}), supports $S$-symmetry with ${\cal S}=\sigma_{z}\otimes\mathbb{1}_{r\times r}$. With ${\cal T}=\mathbb{1}_{r \times r}\otimes(-i\sigma_{y})$ now being the matrix which implements a spin flip, $T$-symmetry is fulfilled if $(-i\sigma_{y})X(i\sigma_{y})\,=\,X^{\ast}$, $X=A,B$. Applying this relation to $A$ and $B$, we get $x_{22}=x_{11}^{\ast}$, $x_{21}=-x_{12}^{\ast}$, $x=a,b$. These constraints replace the stronger condition of real hopping amplitudes imposed by $T$ in the spinless case, resulting in unpinned band degeneracies.
%%%%%%%%%%%%%%%%%%%%%%%%%%%%%
%%%%%%%%%%%%%%%%%%%%     FIGURE 2
%%%%%%%%%%%%%%%%%%%%%%%%%%%%%%%%%%%%%%%%%%%%%%%%%%%%%%%%%%%%%%%%%%%%%%%%%%%%%%%%
\begin{figure}
\includegraphics[scale=0.41]{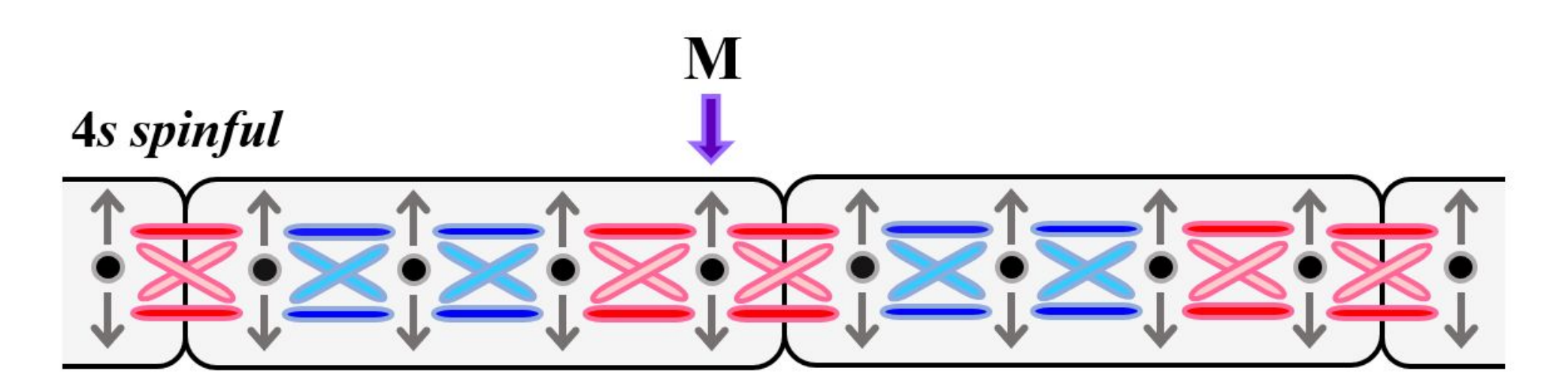}
\caption{(Color online) The $4s$-chain populated by spinful fermions. The colored segments represent bonds with different strengths;
up and down arrows illustrate the spin degree of freedom. A site-mirror point is indicated by M.}
\end{figure}
%%%%%%%%%%%%%%%%%%%%%%%%%%%%%%%%

To see this, let us remove $T$ and consider $\tilde{Q}^{4s}$ in Eq. (\ref{Q4sSpinful}), now with unconstrained $A$ and $B$. For general $A$ and $B$, $\text{det}[\tilde{Q}^{4s}]\,=\,p^{\ast}z^{2}+qz+p$, where
%%%%%%%%%%%%%%%%%%%%%%%%%%%%%%%%
\begin{align}\label{pq}
p\,=\,&\text{det}A\,\text{det}B\,=\,|p|\,e^{i\alpha},\\
\nonumber q\,=\,&2\sum_{x\neq y}\text{Re}(x_{11}x_{22}^{\ast}y_{12}y_{21}^{\ast})\,-\,2\text{Re}(a_{11}a_{22}^{\ast}b_{11}b_{22}^{\ast})\\
\nonumber    -\,&2\text{Re}(a_{12}a_{21}^{\ast}b_{12}b_{21}^{\ast})\,-\,4\sum_{x\neq y}\text{Im}(x_{11}x_{12}^{\ast})\text{Im}(y_{21}y_{22}^{\ast}),
\end{align}
%%%%%%%%%%%%%%%%%%%%%%%%%%%%%%%%%%%%%%%%%%%%
with $x,y=a,b$. The condition $\text{det}[Q(k)]=0$ yielding a zero-energy node is fulfilled if $z\,=\,z_{\pm}\,=\,(t\pm\sqrt{t^{2}-1})e^{i\alpha}$, where $t\equiv-q/(2|p|)$. Since $|z_{\pm}|=1$, one must have $t\in[-1,1]$, in which case $z_{\pm}\,=\,e^{i(\pm\theta+\alpha)}$, with $\theta\,=\,\arctan(\sqrt{1-t^{2}}/t)$ if $0\leq t\leq1$ or $\theta\,=\,\arctan(\sqrt{1-t^{2}}/t)+\pi$ if $-1\leq t<0$. Therefore, a pair of nodes occurs at $k\,=\,k_{\pm}\,=\,\pm\theta+\alpha$ and they move (asymmetrically with respect to $k=0$) as the phases $\theta$ and $\alpha$ change. It follows from the definition of $t$ and Eqs. (\ref{pq}) that satisfying $t \in [-1,1]$ demands fine-tuning the microscopic parameters, meaning that such a node would be accidental. These nodes are shown in a movie appended to the SM \cite{Malard2017}.

We now reintroduce the $T$-constraints $x_{22}=x_{11}^{\ast}$, $x_{21}=-x_{12}^{\ast}$, $x=a,b,$ for which $p$ in Eq. (\ref{pq}) becomes a real positive number, i.e. $\alpha=0$, $|p|=p$. Also, under the $T$-constraints $t\in(-1,1)$, and hence $|z_{\pm}|=1$ {\em with no further constraints on the parameters}. It follows that $z_{\pm}\,=\,e^{\pm i\theta}$ meaning that two nodes occur at the BZ points $k\,=\,k_{\pm}\,=\,\pm\theta$, with $\theta$ as given above but excluding $t=\pm1$. The effect of $T$ is thus to turn the former asymmetric pair of accidental nodes into a symmetric pair of \emph{movable but not removable} degeneracies. Fig. 3 illustrates the spectrum for two parameter configurations, with the parameters $x_{ij}$, $x=a,b$, written as $x_{ij}=|x_{ij}|\text{exp}(i\theta_{x_{ij}})$. At the node for positive (negative) $k$, the two degenerate states have both spin down (up), so the four zero-energy states together form two Kramers pairs. In the SM \cite{Malard2017} the reader will find movies of the spectrum which fully exposes the motion of the nodes in the BZ for different parameter variations.
%%%%%%%%%%%%%%%%%%%%%%%%%%%%%
%%%%%%%%%%%%%%%%%%%%     FIGURES 3-4

%%%%%%%%%%%%%%%%%%%%%%%%%%%%%%%%%%%%%%%%%%%%%%%%%%%%%%%%%%%%%%%%%%%%%%%%%%%%%%%%
\begin{figure}
\includegraphics[scale=0.45]{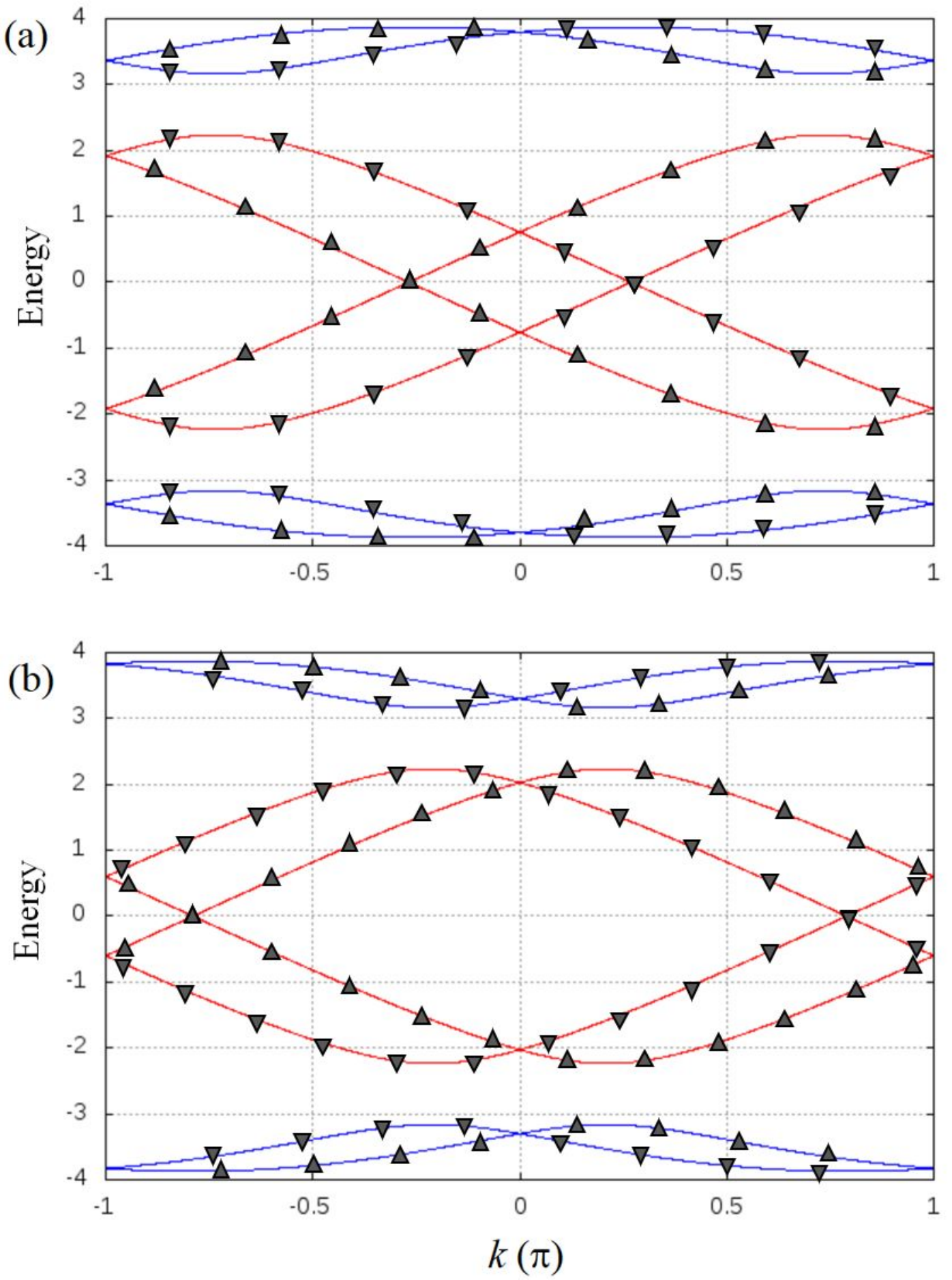}
\caption{(Color online) Spectrum of the spinful $4s$-chain for (a) $\theta_{a_{11}}=1.53\pi$ and (b) $\theta_{a_{11}}=0.04\pi$. For both (a) and (b): $|a_{11}|=2, |a_{12}|=1, |b_{11}|=\sqrt{2}/2, |b_{12}|=\sqrt{2}$ and $\theta_{a_{12}}=\pi/6$, $\theta_{b_{11}}=\pi/3$, $\theta_{b_{12}}=\pi/12$. Color code: Red and blue are employed on the bands to highlight the symmetry of the spectrum around Energy $=0$ and $k=0$, a consequence of chiral and time-reversal symmetries, respectively. Up and down triangles represent the two opposite spin orientations.}
\end{figure}
\begin{figure}
\includegraphics[scale=0.32]{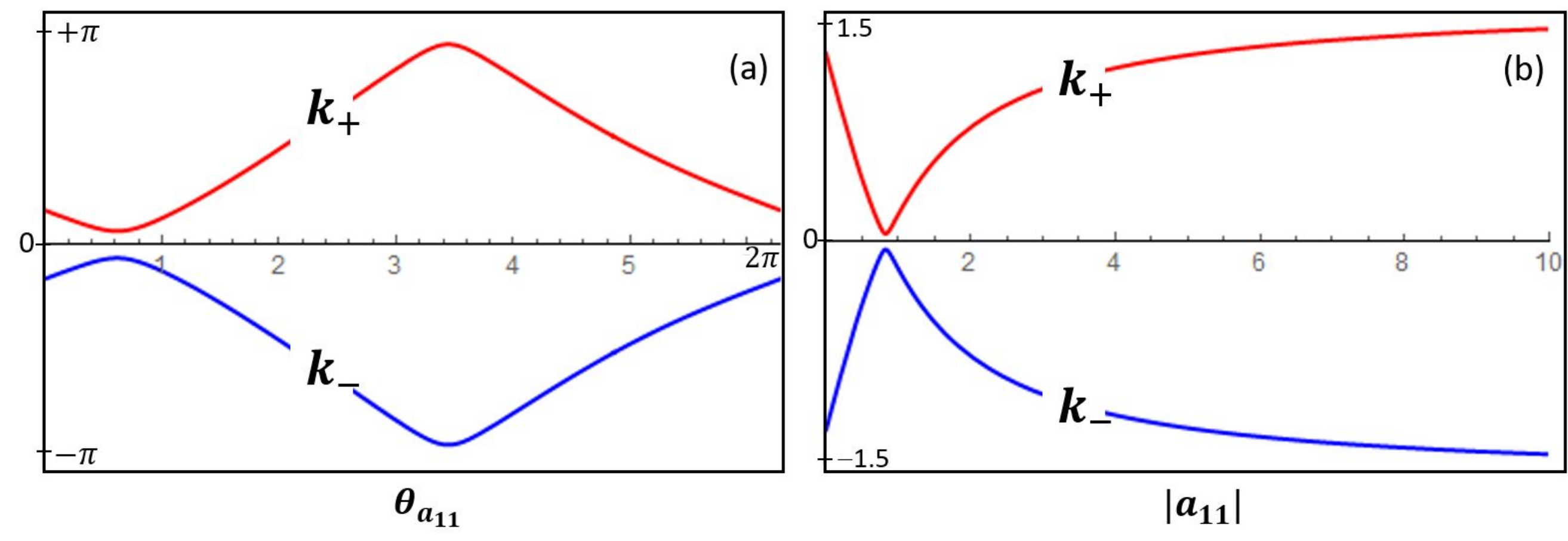}
\caption{(Color online) The locations $k_{\pm}$ of the nodes as a function of (a) $\theta_{a_{11}}$ with $|a_{11}|=1$ and (b) $|a_{11}|$ with $\theta_{a_{11}}=\pi/5$. For both (a) and (b): $|a_{12}|=|b_{11}|=|b_{12}|=1$ and $\theta_{a_{12}}=5\pi/8$, $\theta_{b_{11}}=2\pi/3$, $\theta_{b_{12}}=\pi/4$.}
\end{figure}
%%%%%%%%%%%%%%%%%%%%%%%%%%%%%%%%
The BZ locations of the nodes, given by $k_{\pm}\,=\,\pm\theta$, are shown in Fig. 4 as a function of the phase and of the modulus of $a_{11}$. Fig. 4(a) shows that as $\theta_{a_{11}}$ goes from 0 to $2\pi$, the nodes at opposite sides of the BZ bounce back and forth between the center and the zone boundaries. Varying $|a_{11}|$ causes the nodes to initially approach each other, but they are eventually pushed apart, as shown in Fig. 4(b). In neither case do the nodes merge at $k=0$ or at $k=\pm\pi$. Mathematically, this fact follows from $z_{\pm}$ being complex numbers, hence $z_{\pm}\neq\pm1$ and thus $k_{\pm}\neq0,\pm\pi$. Differently from the Weyl nodes \cite{Murakami2007} and the triple point fermions discussed in Ref. \onlinecite{Fulga2017} (which, in both cases, are topologically protected only locally), our symmetry-enforced nodes cannot coalesce and annihilate. However, the effective repulsion between the nodes as they symmetrically approach the center or the boundaries of the BZ is not easily explained by symmetry alone. Topology may conceivably also play a role, similar to the case of nonsymmorphic degeneracies which come with a global topological invariant \cite{Zhao2016}.

To conclude our analysis, we show that the form of $\tilde{Q}^{4s}$ in Eq. (\ref{Q4sSpinful}) follows from the combination of $T$ and a site-mirror ($M$) symmetries. For that we construct the unitary site-mirror transformation $M(k)=I(k){\cal T}$ formed out of site-inversion $I(k)={\cal \tilde{I}}(k) \times \!\parallel$ times the spin flip ${\cal T}$. For the spinful chain, ${\cal \tilde{I}}(k)\!=\!{\cal I}(k)\!\otimes\!\mathbb{1}_{2\times2}$. Employing Eq. (\ref{Ik}),
%%%%%%%%%%%%%%%%%%%%%%%%%%%%%%%%%%%%%%%%%%%%
\begin{equation}\label{Mk}
{\cal M}(k)\,\equiv\,{\cal \tilde{I}}(k)\,{\cal T}\,=\,\begin{bmatrix}
    \tilde{R}_{1}(k) & 0 \\
    0 & \tilde{R}_{2}(k) \\
\end{bmatrix},
\end{equation}
%%%%%%%%%%%%%%%%%%%%%%%%%%%%%%%%%%%%%%%%%%%%
with $\tilde{R}_{1}(k)\!=\!R_{1}(k)\!\otimes\!(-i\sigma_{y})$ and $ \tilde{R}_{2}(k)\!=\!R_{2}(k)\!\otimes\!(-i\sigma_{y}).$ Demanding that $M(k)$ is a symmetry transformation yields: ${\cal M}(k)\,{\cal H}(-k)\,{\cal M}^{-1}(k)\,=\,{\cal H}(k)$ \cite{Malard2017}. Bringing in Eqs. (\ref{Hk}) and (\ref{Mk}), and using that ${\cal M}^{-1}(k)=-{\cal M}(-k)$, it follows that $\tilde{R}_{1}(k)\, Q(-k)\,\tilde{R}_{2}(-k)\,=\,-Q(k)$. Inserting a generic $Q(k)$ with $r=4$ into this relation leads to: $(-i\sigma_{y})Q_{ij}(i\sigma_{y})\,=\,Q_{jj}$, with $i\neq j=1,2$, and $Q_{ij}$ the $k$-independent $2\times2$ matrices appearing in $Q(k)$. Combining this with the fact that $T$ constrains these matrices as $(-i\sigma_{y})Q_{ij}(i\sigma_{y})\,=\,Q_{ij}^{\ast}$, we get $Q_{ij}\,=\,Q_{jj}^{\ast}$, with $i\neq j=1,2$. This means that $\tilde{Q}^{4s}$ in Eq. (\ref{Q4sSpinful}) is indeed the most general matrix describing a spinful $STM$-invariant chain with $r=4$ sites per unit cell. The analysis can be extended to unit cells with $r>4$, and one concludes that any spinful $STM$-invariant chain exhibits a pair of movable but not removable degeneracies.

\emph{Symmetry classes} -  Let us briefly discuss the symmetry classes of the studied models. In the presence of $S$- and $T$-symmetries, the multiband spinless (spinful) chain belongs to class BDI (CII) \cite{Malard2017} of the Altland-Zirnbauer classification \cite{Chiu2016}. This means that the spinless $STI$- and the spinful $STM$-invariant chains are at the boundary between trivial and topological insulating phases which, in both cases, can be characterized by a $Z$-winding number. Breaking the $I$ or $M$-symmetry will generically open a gap at zero energy, driving the system into either one of the insulating regimes. This is similar to how a nonsymmorphic symmetry correlates with a topological phase transition in models of 2D Dirac semimetals \cite{Young2015}.

\emph{Summary} - We have identified a class of 1D electronic tight-binding models which allow the presence of spin-orbit interactions and whose band structures exhibit \emph{movable but not removable} degeneracies without the presence of a nonsymmorphic symmetry. Chiral, time-reversal and site-mirror symmetry comprise a sufficient set of symmetries for the emergence of this type of degeneracy which, in the case at hand, come in the form of a Kramers related pair of nodes. An interesting open problem is whether these nodes are endowed with a global topological invariant, analogous to the case of nonsymmorphic degeneracies \cite{Zhao2016}. Possible generalizations include adding longer-range odd-neighbor hoppings or superconducting pairing that preserve the enforcing symmetries. Of obvious interest would be to extend our finding to higher dimensions. This could open a pathway to search for new nodal semimetals in symmorphic crystals and, important for applications, in the presence of strong spin-orbit interactions.

\emph{Acknowledgements} -  It is a pleasure to thank Jens Bardarson, Jan Budich, Maria Hermanns, Gia Japaridze, Andreas Schnyder, Alexander Stolin, and Long Zhang for valuable discussions.

This research was supported by the Swedish Research Council (Grant No. 621-2014-5972).

\begin{widetext}
\setcounter{figure}{0}
\setcounter{equation}{0}
\setcounter{section}{0}
\newpage
\section{\large Supplemental Material}
\renewcommand\thefigure{S\arabic{figure}}
\renewcommand\theequation{S\arabic{equation}}

\vskip 0.5 cm

\section{Model}

\subsection{Lattice Hamiltonian}

In this Section we derive the Bloch matrix for the considered class of models - Eq. (1) in the accompanying Letter [SM1] - starting from a tight-binding lattice Hamiltonian in position space. We consider a translational invariant one-dimensional (1D) lattice with $N$ unit cells, each cell containing $r\in2\mathds{N}$ sites, and populated by spinless or spinful fermions with nearest-neighbor hopping. The Hamiltonian $H$ can be written as a sum of intra-cell and inter-cell terms,
%%%%%%%%%%%%%%%%%%%%%%%%%%%%%%%%%%%%%%%%%%%%
\begin{equation} \label{H}
H\,=\,\sum^{N}_{m=1}\left[\sum^{r-1}_{n=1}\sum_{\tau=\pm}{\cal H}_{\text{intra}}\,+\,\sum_{\tau=\pm}{\cal H}_{\text{inter}}\right]\,+\,\mbox{H.c.},
\end{equation}
%%%%%%%%%%%%%%%%%%%%%%%%%%%%%%%%%%%%%%%%%%%%
with
%%%%%%%%%%%%%%%%%%%%%%%%%%%%%%%%%%%%%%%%%%%%
\begin{equation} \label{Hintra}
{\cal H}_{\text{intra}}\,=\,\alpha_{n}^{\tau}\,c_{m,n}^{\tau\,\dagger}c_{m,n+1}^{\tau}\,+\,\beta_{n}^{\tau}\,c_{m,n}^{\tau\,\dagger}c_{m,n+1}^{-\tau},
\end{equation}
\begin{equation} \label{Hinter}
{\cal H}_{\text{inter}}\,=\,\alpha_{r}^{\tau}\,c_{m,r}^{\tau\,\dagger}c_{m+1,1}^{\tau}\,+\,\beta_{r}^{\tau}\,c_{m,r}^{\tau\,\dagger}c_{m+1,1}^{-\tau},
\end{equation}
%%%%%%%%%%%%%%%%%%%%%%%%%%%%%%%%%%%%%%%%%%%%
where $c_{m,n}^{\tau\,\dagger}$ ($c_{m,n}^{\tau}$) creates (annihilates) a particle at site $n$ in unit cell $m$ with spin projection $\tau=\pm$ in the case of spinful fermions; if spinless fermions, the $\tau$-index is dropped. The complex scalars $\alpha_{n}^{\tau}$ ($\beta_{n}^{\tau}$) are amplitudes for spin-conserving (-flipping) hopping. In the case of spinless fermions, $\beta_{n}=0$.

\subsection{Momentum space Hamiltonian}

By Fourier transforming,
%%%%%%%%%%%%%%%%%%%%%%%%%%%%%%%%%%%%%%%%%%%%
\begin{equation} \label{FourierTransform}
c_{m,n}^{\tau}\,=\,\frac{1}{\sqrt{N}}\sum^{\pi}_{k=-\pi}\,c_{k,n}^{\tau}e^{ikm},
\end{equation}
%%%%%%%%%%%%%%%%%%%%%%%%%%%%%%%%%%%%%%%%%%%%
where $k=k_{j}=\pm2\pi j/N$, $j=0,1,...,N/2$, the Hamiltonian defined by Eqs. (\ref{H})-(\ref{Hinter}) turns into
%%%%%%%%%%%%%%%%%%%%%%%%%%%%%%%%%%%%%%%%%%%%
\begin{equation} \label{Hfouriertransformed}
H\,=\,\sum^{\pi}_{k=-\pi}\,\sum^{r}_{n,n'=1}\,\sum_{\tau,\tau'=\pm}\,c_{k,n}^{\tau\dagger}\,{\cal H}_{n\tau,n'\tau'}(k)\,c_{k,n'}^{\tau'}
\end{equation}
%%%%%%%%%%%%%%%%%%%%%%%%%%%%%%%%%%%%%%%%%%%%
with
%%%%%%%%%%%%%%%%%%%%%%%%%%%%%%%%%%%%%%%%%%%%
\begin{align} \label{MatrixElementsofH}
\nonumber {\cal H}_{n\tau,n'\tau'}(k)\,=\,&\alpha_{n}^{\tau}\delta_{n',n+1}\delta_{\tau',\tau}\,+\,\beta_{n}^{\tau}\delta_{n',n+1}\delta_{\tau',-\tau}\,+\\
\nonumber +\,&\alpha_{n-1}^{\tau\,\ast}\delta_{n',n-1}\delta_{\tau',\tau}\,+\,\beta_{n-1}^{\tau\,\ast}\delta_{n',n-1}\delta_{\tau',-\tau}\,+\\
\nonumber+\,&\alpha_{r}^{\tau\,\ast}e^{-ik}\delta_{n,1}\delta_{n',r}\delta_{\tau',\tau}\,+\,\beta_{r}^{\tau\,\ast}e^{-ik}\delta_{n,1}\delta_{n',r}\delta_{\tau',-\tau}\,+\\
+\,&\alpha_{r}^{\tau}e^{+ik}\delta_{n,r}\delta_{n',1}\delta_{\tau',\tau}\,+\,\beta_{r}^{\tau}e^{+ik}\delta_{n,r}\delta_{n',1}\delta_{\tau',-\tau}.
\end{align}
%%%%%%%%%%%%%%%%%%%%%%%%%%%%%%%%%%%%%%%%%%%%

Again, note that the $\tau$-index in Eqs. (\ref{Hfouriertransformed})-(\ref{MatrixElementsofH}) is absent in the case of spinless fermions.

\subsection{Bloch matrix}

We partition the lattice into two sublattices, one formed out of the odd-labelled intra-cell sites and the other from the even-labelled ones. For spinless fermions, we define an $r$-dimensional row spinor $c^{\dagger}_{k}$, ordering the creation operators in the following way:
%%%%%%%%%%%%%%%%%%%%%%%%%%%%%%%%%%%%%%%%%%%%
\begin{equation} \label{ChiralSpinor}
c^{\dagger}_{k}\,=\,(c_{k,1}^{\dagger},...,c_{k,r-1}^{\dagger},c_{k,2}^{\dagger},...,c_{k,r}^{\dagger}),
\end{equation}
%%%%%%%%%%%%%%%%%%%%%%%%%%%%%%%%%%%%%%%%%%%%
where the first (last) $r/2$ entries are creation operators defined on the sublattice of odd (even) intra-cell sites. An $r$-dimensional column spinor $c_{k}$ is defined by ordering the annihilation operators in the same way. For spinful fermions, we define $2r$-dimensional spinors also in terms of sublattice blocks, with the operators for opposite spins at the same intra-cell site placed next to each other.

In the above spinor representation, the Hamiltonian (\ref{Hfouriertransformed})-(\ref{MatrixElementsofH}) is written as
%%%%%%%%%%%%%%%%%%%%%%%%%%%%%%%%%%%%%%%%%%%%
\begin{equation} \label{FourierTransformed2}
H=\sum_{k}\,c^{\dag}_{k}\,{\cal H}(k)\,c_{k},
\end{equation}
%%%%%%%%%%%%%%%%%%%%%%%%%%%%%%%%%%%%%%%%%%%%
with the $r\times r$ ($2r\times 2r$) Bloch matrix ${\cal H}(k)$ for spinless (spinful) fermions given by
%%%%%%%%%%%%%%%%%%%%%%%%%%%%%%%%%%%%%%%%%%%%
\begin{equation} \label{HkAlgebraic}
{\cal H}(k)\,=\,\frac{\sigma_{x}}{2}\otimes[Q(k)+Q^{\dagger}(k)]\,+\,\frac{i\sigma_{y}}{2}\otimes[Q(k)-Q^{\dagger}(k)]
\end{equation}
%%%%%%%%%%%%%%%%%%%%%%%%%%%%%%%%%%%%%%%%%%%%
or, as it appears in Eq. (1) of [SM1],
%%%%%%%%%%%%%%%%%%%%%%%%%%%%%%%%%%%%%%%%%%%%
\begin{equation} \label{HkSM}
{\cal H}(k)\,=\, \begin{bmatrix}
    0 & Q(k) \\
    Q^{\dagger}(k) & 0 \\
\end{bmatrix}.
\end{equation}
%%%%%%%%%%%%%%%%%%%%%%%%%%%%%%%%%%%%%%%%%%%%
In the case of spinless fermions, $Q(k)$ is the $r/2\times r/2$ matrix
%%%%%%%%%%%%%%%%%%%%%%%%%%%%%%%%%%%%%%%%%%%%
\begin{equation} \label{Qk}
Q(k)\,=\, \begin{bmatrix}
    \alpha_{1} & 0 & 0 & \dots & 0 & \alpha^{\ast}_{r}z\\
    \alpha^{\ast}_{2} & \alpha_{3} & 0 & \dots & 0 & 0 \\
    \vdots & \vdots & \vdots & \dots & \vdots & \vdots \\
    0 & 0 & 0 & \dots & \alpha^{\ast}_{r-2} & \alpha_{r-1} \\
\end{bmatrix},
\end{equation}
%%%%%%%%%%%%%%%%%%%%%%%%%%%%%%%%%%%%%%%%%%%%
with $z=e^{-ik}$. For spinful fermions, $Q(k)$ is an $r\times r$ matrix obtained by replacing the amplitudes $\alpha_{n}$ from Eq. (\ref{Qk}) by the $2\times 2$ matrices
%%%%%%%%%%%%%%%%%%%%%%%%%%%%%%%%%%%%%%%%%%%%
\begin{equation} \label{A}
A_{n}\,=\, \begin{bmatrix}
    \alpha_{n}^{+} & \beta_{n}^{-}\\
    \beta_{n}^{+} & \alpha_{n}^{-}
\end{bmatrix}.
\end{equation}
%%%%%%%%%%%%%%%%%%%%%%%%%%%%%%%%%%%%%%%%%%%%

One sees that, in the spinor representation (\ref{ChiralSpinor}), the Bloch matrix (\ref{HkSM}) assumes an off-diagonal form. This is a consequence of the lack of matrix elements within the same sublattice, a property of chiral (or sublattice) symmetry [SM2]. We will refer to Eq. (\ref{ChiralSpinor}) (and its extension to the spinful case) as the ``chiral representation".

\section{Discrete symmetries and symmetry classes}

A system has chiral ($S$), time reversal ($T$) and particle-hole ($C$) symmetries if its Bloch matrix ${\cal H}(k)$ satisfies the following invariance relations:
%%%%%%%%%%%%%%%%%%%%%%%%%%%%%%%%%%%%%%%%%%%%
\begin{equation} \label{ChiralSymmetry}
{\cal S}\,{\cal H}(k)\,{\cal S}^{-1}\,=\,-{\cal H}(k),
\end{equation}
\begin{equation} \label{TimeReversalSymmetry}
{\cal T}\,{\cal H}(k)\,{\cal T}^{-1}\,=\,{\cal H}^{\ast}(-k),
\end{equation}
\begin{equation} \label{ParticleHoleSymmetry}
{\cal C}\,{\cal H}(k)\,{\cal C}^{-1}\,=\,-{\cal H}^{\ast}(-k),
\end{equation}
%%%%%%%%%%%%%%%%%%%%%%%%%%%%%%%%%%%%%%%%%%%%
where ${\cal S}$, ${\cal T}$ and ${\cal C}$ are matrices representing, in a chosen spinor representation, chiral, time reversal and particle-hole transformations, respectively [SM3].

Considering the class of models above, Eqs. (\ref{HkAlgebraic})-(\ref{Qk}), one defines the chiral transformation as ${\cal S}={\cal P}_{\text{o}}-{\cal P}_{\text{e}}$, where ${\cal P}_{\text{o}}$ (${\cal P}_{\text{e}}$) is the projector onto the sublattice of odd (even) intra-cell sites. With this definition, ${\cal S}$ is given, in the chiral representation (\ref{ChiralSpinor}) (and its extension to the spinful case), by ${\cal S}\,=\,\sigma_{z}\otimes1\!\!1_{n\times n}$, with $n=r/2\,(r)$ for spinless (spinful) fermions, and thus ${\cal S}^{2}\,=\,1\!\!1$. One easily checks that Eq. (\ref{ChiralSymmetry}) is satisfied with ${\cal H}(k)$ given by Eq. (\ref{HkAlgebraic}).

In the case of spinless fermions, ${\cal T}\,=\,1\!\!1_{r\times r}$. Eq. (\ref{TimeReversalSymmetry}), with ${\cal H}(k)$ given by Eqs. (\ref{HkSM})-(\ref{Qk}), then implies that $\alpha_{n}\in\mathds{R}$. If the particles are spinful, ${\cal T}$ is a spin flip operation which, in the chiral representation, reads ${\cal T}\,=\,1\!\!1_{r\times r}\otimes(-i\sigma_{y})$. In this case, the symmetry relation (\ref{TimeReversalSymmetry}) implies: $(-i\sigma_{y})A_{n}(i\sigma_{y})\,=\,A^{\ast}_{n}$. Also, ${\cal T}^{2}\,=\,\pm1\!\!1$, with the plus (minus) sign applying to spinless (spinful) fermions.

As for particle-hole symmetry, since ${\cal S}\,=\,{\cal T}{\cal C}$ [SM3], it follows that ${\cal C}\,=\,\pm{\cal T}{\cal S}$, with the plus (minus) sign applying for spinless (spinful) fermions. One can check that, having fulfilled equalities (\ref{ChiralSymmetry})-(\ref{TimeReversalSymmetry}), Eq. (\ref{ParticleHoleSymmetry}) for particle-hole symmetry is satisfied automatically with ${\cal C}\,=\,\pm{\cal T}{\cal S}$. It follows that, in the chiral representation, ${\cal C}\,=\,\sigma_{z}\otimes1\!\!1_{r/2\times r/2}$ in the case of spinless fermions and ${\cal C}\,=\,\sigma_{z}\otimes1\!\!1_{r/2\times r/2}\otimes(i\sigma_{y})$ if the fermions are spinful. As a consequence, ${\cal C}^{2}\,=\,\pm1\!\!1$, with the plus (minus) sign applying to spinless (spinful) fermions.

Given the above possibilities for ${\cal S}^{2}$, ${\cal T}^{2}$ and ${\cal C}^{2}$, we conclude that while the $STC$-invariant spinless realizations of this class of models belong to symmetry class BDI of the Altland-Zirnbauer classification, the spinful cases are in class CII, with the gapped phases of both classes being distinguished by a $Z$-winding number [SM3].

\section{Site-inversion and site-mirror transformations}

In this Section, we derive the operators that implement site-inversion and site-mirror transformations on a spinless and on a spinful chain, respectively.

Starting with site-inversion, let us consider the minimal spinless $4s$-chain introduced in [SM1] with the inversion point located at site 4 of the $m=0$ unit cell, as shown in Fig. S1. One can see that such a site-inversion transformation acts on the lattice operators according to:
%%%%%%%%%%%%%%%%%%%%%%%%%%%%%%%%%%%%%%%%%%%%
\begin{equation}
\nonumber c_{m}^{1,2,3} \rightarrow c_{-m+1}^{3,2,1}, \qquad c_{m}^{4} \rightarrow c_{-m}^{4}.
\end{equation}
%%%%%%%%%%%%%%%%%%%%%%%%%%%%%%%%%%%%%%%%%%%%
%%%%%%%%%%%%%%%%%%%%%%%%%%%%%
%%%%%%%%%%%%%%%%%%%%     FIGURE 1
%%%%%%%%%%%%%%%%%%%%%%%%%%%%%%%%%%%%%%%%%%%%%%%%%%%%%%%%%%%%%%%%%%%%%%%%%%%%%%%%
\begin{figure}
\includegraphics[scale=0.45]{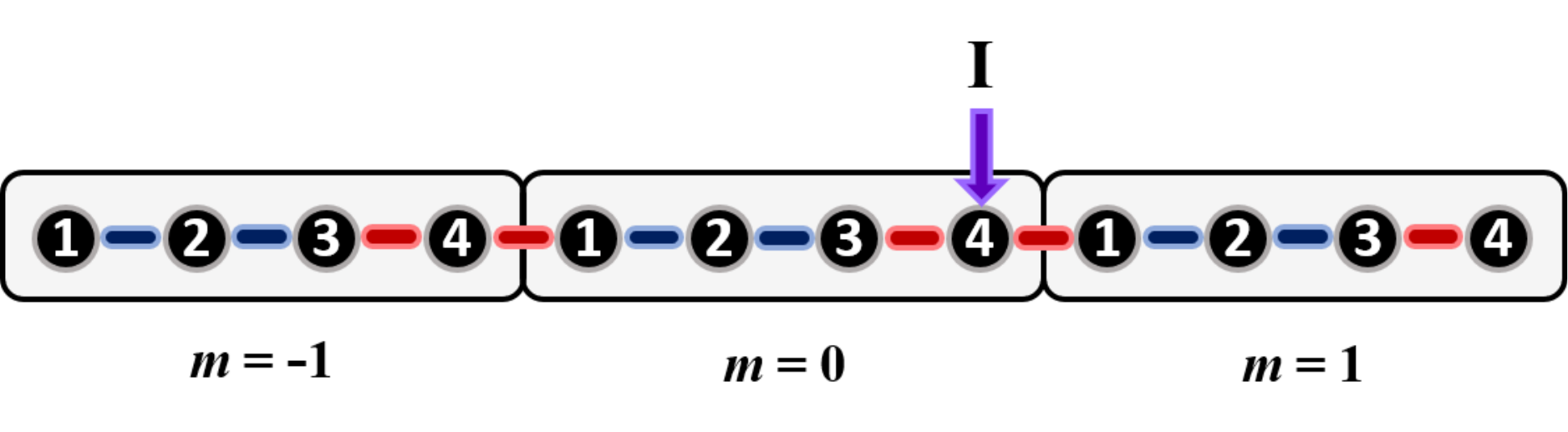}
\caption{(Color online) Spinless $4s$-chain with a site-inversion point located at site 4 of the $m=0$ unit cell.}
\end{figure}
%%%%%%%%%%%%%%%%%%%%%%%%%%%%%%%%

Fourier transforming the above relations (by using Eq. (\ref{FourierTransform}), with the spin index $\tau$ removed) one finds that the site-inversion acts on the momentum-space spinors as:
%%%%%%%%%%%%%%%%%%%%%%%%%%%%%%%%%%%%%%%%%%%%
\begin{equation}\label{TransfChiralSpinor}
c_{k}\,=\,\begin{bmatrix}
    c_{k}^{1}\\
    c_{k}^{3}\\
    c_{k}^{2}\\
    c_{k}^{4}
\end{bmatrix} \rightarrow \tilde{c}_{k}\,=\,z\begin{bmatrix}
    c_{-k}^{3}\\
    c_{-k}^{1}\\
    c_{-k}^{2}\\
    z^{\ast}c_{-k}^{4}
\end{bmatrix} ,
\end{equation}
%%%%%%%%%%%%%%%%%%%%%%%%%%%%%%%%%%%%%%%%%%%%
where $z=e^{-ik}$.

Consider the unitary operator $I(k)\,=\,{\cal I}(k)\times\!\parallel$, where the ``hard wall" operator $\parallel$ reverses momentum and
%%%%%%%%%%%%%%%%%%%%%%%%%%%%%%%%%%%%%%%%%%%%
\begin{equation}
\nonumber {\cal I}(k)\,=\,z\begin{bmatrix}
    0 & 1 & 0 & 0 \\
    1 & 0 & 0 & 0 \\
    0 & 0 & 1 & 0 \\
    0 & 0 & 0 & z^{\ast}
\end{bmatrix},
\end{equation}
or, equivalently,
\begin{equation} \label{IkSM}
{\cal I}(k)\,=\, \begin{bmatrix}
    R_{1}(k) & 0 \\
    0 & R_{2}(k) \\
\end{bmatrix},
\end{equation}
\begin{equation}\label{R1R2}
R_{1}(k)\,=\,z\,\text{adiag}(1\,1),\,\,R_{2}(k)\,=\,z\,\text{diag}(1\,z^{\ast}),
\end{equation}
%%%%%%%%%%%%%%%%%%%%%%%%%%%%%%%%%%%%%%%%%%%%
with the symbol $\text{diag}$ ($\text{adiag}$) denoting a diagonal (anti-diagonal) matrix. One can easily check that $I(k)$ is the operator which performs the site-inversion transformation (\ref{TransfChiralSpinor}), i.e. $I(k)\,c_{k}\,=\,\tilde{c}_{k}$.

By applying the same procedure that led to Eq. (\ref{TransfChiralSpinor}) for $r=4$, one arrives at the spinor transformation rule for an arbitrarily sized unit cell with $r>4$. It is then found that the corresponding site-inversion operator is defined in the same way, i.e. $I(k)\,=\,{\cal I}(k)\times\!\parallel$, with ${\cal I}(k)$ of the same form as in Eq. (\ref{IkSM}) (since the transformation acts only within one sublattice), however now with the matrices in Eq. (\ref{R1R2}) replaced by new $r/2\times r/2$ matrices $R_{1}(k)$ and $R_{2}(k)$.

Since inversion is defined as a transformation on position space only (it does not act on spin space), the unitary site-inversion operator for the spinful chain reads $\tilde{I}(k)\,=\,{\cal \tilde{I}}(k)\times\!\parallel$, with ${\cal \tilde{I}}(k)\,=\,{\cal I}(k)\otimes1\!\!1_{2\times2}$, the same ${\cal I}(k)$ as in Eq. (\ref{IkSM}) and the corresponding $R_{1}(k)$ and $R_{2}(k)$. In the case of spinful fermions, we need also the unitary operator implementing a site-mirror transformation: $M(k)\,=\,\tilde{I}(k)\,\cal T$, formed out of site-inversion $\tilde{I}(k)$ times the spin flip ${\cal T}\,=\,1\!\!1_{r\times r}\otimes(-i\sigma_{y})$. We define ${\cal M}(k)\,\equiv\,{\cal \tilde{I}}(k)\,{\cal T}$ which, through substitution of ${\cal \tilde{I}}(k)\,=\,{\cal I}(k)\otimes1\!\!1_{2\times2}$ and of Eq. (\ref{IkSM}), writes
%%%%%%%%%%%%%%%%%%%%%%%%%%%%%%%%%%%%%%%%%%%%
\begin{equation}\label{MkSM}
{\cal M}(k)\,=\,\begin{bmatrix}
    \tilde{R}_{1}(k) & 0 \\
    0 & \tilde{R}_{2}(k) \\
\end{bmatrix},
\end{equation}
%%%%%%%%%%%%%%%%%%%%%%%%%%%%%%%%%%%%%%%%%%%%
with
%%%%%%%%%%%%%%%%%%%%%%%%%%%%%%%%%%%%%%%%%%%%
\begin{equation}\label{R1R2tilde}
\tilde{R}_{1}(k)\,=\,R_{1}(k)\otimes(-i\sigma_{y}),\,\,\tilde{R}_{2}(k)\,=\,R_{2}(k)\otimes(-i\sigma_{y}).
\end{equation}
%%%%%%%%%%%%%%%%%%%%%%%%%%%%%%%%%%%%%%%%%%%%

\section{Spinless chain with chiral, time reversal and site-inversion symmetries}

As established above, the class of models (\ref{HkSM})-(\ref{Qk}) have built-in $S$-symmetry and, for spinless fermions with real hopping amplitudes, also $T$-symmetry. Additionally, we now impose site-inversion ($I$) symmetry:
%%%%%%%%%%%%%%%%%%%%%%%%%%%%%%%%%%%%%%%%%%%%
\begin{equation}\label{SiteInversionSymmetry}
{\cal I}(k)\,{\cal H}(-k)\,{\cal I}^{-1}(k)\,=\,{\cal H}(k).
\end{equation}
%%%%%%%%%%%%%%%%%%%%%%%%%%%%%%%%%%%%%%%%%%%%

Inserting Eqs. (\ref{HkSM}) and (\ref{IkSM}) into Eq. (\ref{SiteInversionSymmetry}) and using that ${\cal I}^{-1}(k)={\cal I}(-k)$, we get an invariance relation for the matrix $Q(k)$:
%%%%%%%%%%%%%%%%%%%%%%%%%%%%%%%%%%%%%%%%%%%%
\begin{equation}\label{invarianceQR1R2}
R_{1}(k)\,Q(-k)\,R_{2}(-k)\,=\,Q(k).
\end{equation}
%%%%%%%%%%%%%%%%%%%%%%%%%%%%%%%%%%%%%%%%%%%%

Finally, we subject $Q(k)$ from Eq. (\ref{Qk}) to Eq. (\ref{invarianceQR1R2}), pulling out the constraints imposed by $I$-symmetry on the parameters $\alpha_{n}$. Again taking as an example the case with $r=4$ for which Eqs. (\ref{R1R2}) apply, we get a $Q(k)$ of the form:
%%%%%%%%%%%%%%%%%%%%%%%%%%%%%%%%%%%%%%%%%%%%
\begin{equation} \label{Q4sSpinless}
Q(k)\,=\,\begin{bmatrix}
    \alpha_{1} & \alpha_{3}z \\
    \alpha_{1} & \alpha_{3} \\
\end{bmatrix},
\end{equation}
%%%%%%%%%%%%%%%%%%%%%%%%%%%%%%%%%%%%%%%%%%%%
with $\alpha_{1}$ and $\alpha_{3}$ real parameters due to $T$-symmetry. This shows that the matrix $Q^{4s}$ in Eq. (2) of [SM1] is the most general one supporting combined $STI$-symmetries in a spinless chain with $r=4$ sites per unit cell. In general, the procedure leads to the corresponding $STI$-invariant $Q^{rs}$ matrix for any $r>4$.

\section{Spinful chain with chiral, time reversal and site-mirror symmetries}

As we have seen, for the spinful chain $T$-symmetry implies that the $2\times2$ matrices introduced in Eq. (\ref{A}) must satisfy: $(-i\sigma_{y})A_{n}(i\sigma_{y})\,=\,A^{\ast}_{n}$. Let us now add site-mirror ($M$) symmetry:
%%%%%%%%%%%%%%%%%%%%%%%%%%%%%%%%%%%%%%%%%%%%
\begin{equation}\label{SiteMirrorSymmetry}
{\cal M}(k)\,{\cal H}(-k)\,{\cal M}^{-1}(k)\,=\,{\cal H}(k).
\end{equation}
%%%%%%%%%%%%%%%%%%%%%%%%%%%%%%%%%%%%%%%%%%%%

Substituting Eqs. (\ref{HkSM}) and (\ref{MkSM}) into Eq. (\ref{SiteMirrorSymmetry}) and using that ${\cal M}^{-1}(k)=-{\cal M}(-k)$, we receive the invariance relation for $Q(k)$:
%%%%%%%%%%%%%%%%%%%%%%%%%%%%%%%%%%%%%%%%%%%%
\begin{equation}\label{invarianceQR1R2tilde}
\tilde{R}_{1}(k)\,Q(-k)\,\tilde{R}_{2}(-k)\,=\,-Q(k).
\end{equation}
%%%%%%%%%%%%%%%%%%%%%%%%%%%%%%%%%%%%%%%%%%%%

We now get $Q(k)$ from Eq. (\ref{Qk}) with the replacement $\alpha_{n}\rightarrow A_{n}$ and, together with Eq. (\ref{R1R2tilde}), we enforce the symmetry constraint (\ref{invarianceQR1R2tilde}). Again we show the case $r=4$ for which Eq. (\ref{invarianceQR1R2tilde}) can be encoded in the identity: $(-i\sigma_{y})A_{2n-1}(i\sigma_{y})\,=\,A_{2n}$, $n=1,2$. Combining this with $(-i\sigma_{y})A_{n}(i\sigma_{y})\,=\,A^{\ast}_{n}$, we get that $A_{2n-1}^{\ast}\,=\,A_{2n}$, $n=1,2$. Or, in matrix form:
%%%%%%%%%%%%%%%%%%%%%%%%%%%%%%%%%%%%%%%%%%%%
\begin{equation} \label{Q4sSpinfulSM}
Q(k)\,=\,\begin{bmatrix}
    A_{1} & A_{3}^{\ast}z \\
    A_{1}^{\ast} & A_{3} \\
\end{bmatrix},
\end{equation}
%%%%%%%%%%%%%%%%%%%%%%%%%%%%%%%%%%%%%%%%%%%%
where the remaining matrices still have to obey $(-i\sigma_{y})A_{i}(i\sigma_{y})\,=\,A^{\ast}_{i}$, $i=1,3$. With these matrices given by Eq. (\ref{A}), this last constraint translates as: $\alpha_{i}^{-}\,=\,\alpha_{i}^{+\ast}$, $\beta_{i}^{-}\,=\,-\beta_{i}^{+\ast}$, $i=1,3$. We conclude that the matrix $\tilde{Q}^{4s}$ in Eq. (4) of [SM1] (supplemented with the previous constraints in the inner structure of the $2\times2$ matrices) is indeed the most general one describing a spinful $STM$-invariant chain with $r=4$ sites per unit cell. As argued before, repeating the procedure one can obtain the sequence of $STM$-invariant $\tilde{Q}^{rs}$ matrices with $r>4$.

\section{Example: The $6b$- and $6s$-chains}

Adding to the analysis carried out in the accompanying Letter [SM1] for chains with $r=4$ sites per unit cell, here we present the $6b$- and $6s$-chains with $r=6$ sites per unit cell and which are invariant under bond- and site-inversion, respectively. These chains are illustrated in Fig. S2.
%%%%%%%%%%%%%%%%%%%%%%%%%%%%%
%%%%%%%%%%%%%%%%%%%%     FIGURE 2
%%%%%%%%%%%%%%%%%%%%%%%%%%%%%%%%%%%%%%%%%%%%%%%%%%%%%%%%%%%%%%%%%%%%%%%%%%%%%%%%
\begin{figure}
\includegraphics[scale=0.25]{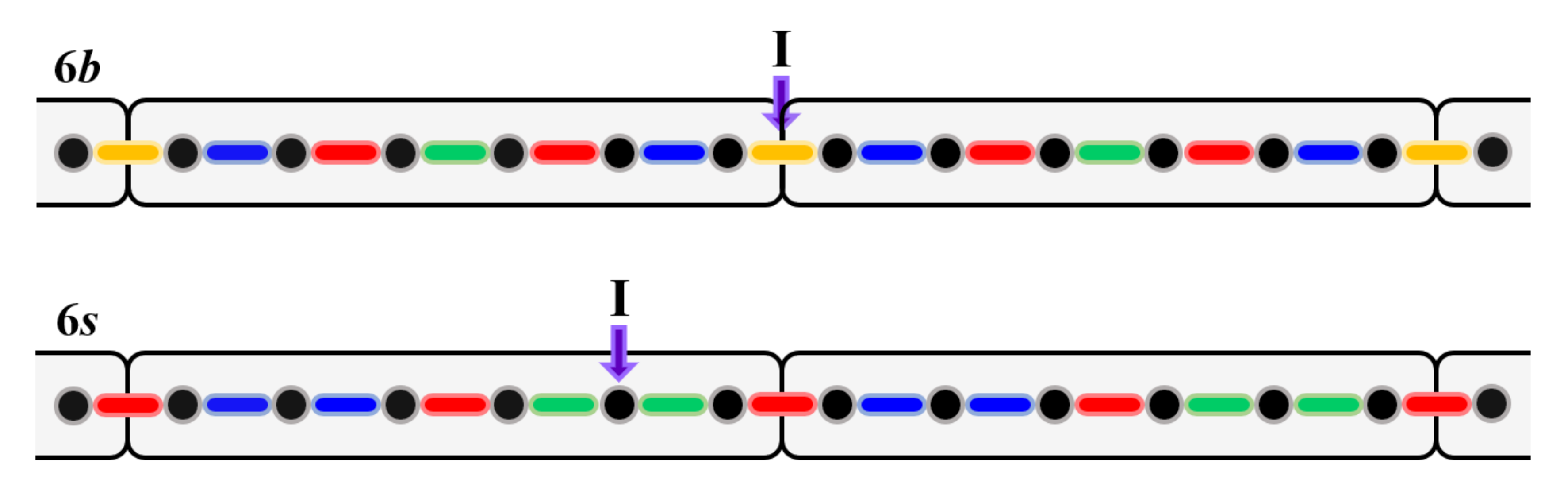}
\caption{(Color online) Chains with bond-inversion symmetry - $6b$ - and with site-inversion symmetry - $6s$. The colored segments represent bonds with different strengths. The inversion point is indicated by I.}
\end{figure}
%%%%%%%%%%%%%%%%%%%%%%%%%%%%%%%%

The $Q$-matrices for the chains in Fig. S2 can be read off from Eq. (\ref{Qk}), with the parameters $\alpha_{n}$, $n=1,2,...,6$, taking on four (three) different values in the $6b$-chain ($6s$-chain) (see Fig. S2). The matrices are thus
%%%%%%%%%%%%%%%%%%%%%%%%%%%%%%%%%%%%%%%%%%%%
\begin{equation} \label{Q6}
Q^{6b}\,=\, \begin{bmatrix}
    a & 0 & dz \\
    b & c & 0 \\
    0 & b & a \\
\end{bmatrix},\qquad\quad Q^{6s}\,=\, \begin{bmatrix}
    a & 0 & bz \\
    a & b & 0 \\
    0 & c & c \\
\end{bmatrix},
\end{equation}
%%%%%%%%%%%%%%%%%%%%%%%%%%%%%%%%%%%%%%%%%%%%
where $z=e^{-ik}$ and $a$, $b$, $c$, $d$ are the hopping amplitudes along the blue, red, green and yellow bonds in Fig. S2, respectively. Due to the assumed $T$-symmetry, $a$, $b$, $c$, $d$ are real numbers which, without loss of generality, we take to be positive.

Alternatively, the form of $Q^{6s}$ could have been obtained from applying the $T$-symmetry condition of real hopping amplitudes together with the $I$-symmetry condition (\ref{invarianceQR1R2}), with
%%%%%%%%%%%%%%%%%%%%%%%%%%%%%%%%%%%%%%%%%%%%
\begin{equation}\label{R1R26sites}
R_{1}(k)\,=\,\begin{bmatrix}
    0 & z & 0 \\
    z & 0 & 0 \\
    0 & 0 & 1 \\
\end{bmatrix}, \qquad\quad R_{2}(k)\,=\,\begin{bmatrix}
    z & 0 & 0 \\
    0 & 0 & 1 \\
    0 & 1 & 0 \\
\end{bmatrix}
\end{equation}
%%%%%%%%%%%%%%%%%%%%%%%%%%%%%%%%%%%%%%%%%%%%
the appropriate matrices when $r=6$.

The condition for the existence of a zero-energy node in the spectrum is: $\text{det}[Q(k)]=0$, subject to $|z|=1$. Applying this condition to the matrices $Q(k)$ given in (\ref{Q6}) gives in each case: $z=z^{6b}=-a^{2}c/(b^{2}d)$ provided that $a^{2}c=b^{2}d$; $z=z^{6s}=-1$ for any $a$, $b$, $c$. Since $z=e^{-ik}$, in both cases the node is pinned at $k=\pm\pi$, i.e. at the boundaries of the BZ. As in the $r=4$ case presented in [SM1], the existence of a bond-inversion solution relies on tuning the model parameters to a certain condition (here $a^{2}c=b^{2}d$), making the node accidental. Differently, site-inversion invariance produces a symmetry-enforced node by constraining the matrix $Q(k)$ in such a way that a nodal solution exists in the whole parameter space of the model. As shown in [SM1], adding spin has the striking effect of unpinning the node (and also of splitting it into a pair of Kramers related nodes), without disrupting the symmetry enforcement.

\section{Band structure}

Using a numerical exact diagonalization method, we obtain the spectrum of a spinful $S$-invariant chain described by the Bloch matrix (\ref{HkSM}) with $Q(k)$ given by Eq. (\ref{Q4sSpinfulSM}), where here $A_{1}$ and $A_{3}$ are taken to be general $2\times2$ matrices \emph{not} constrained by the $T$-symmetry condition $(-i\sigma_{y})A_{n}(i\sigma_{y})\,=\,A^{\ast}_{n}$. Incidentally, breaking $T$-symmetry in this way implies that $(-i\sigma_{y})A_{2n-1}(i\sigma_{y})\,\neq\,A_{2n}$, $n=1,2$, i.e. it simultaneously breaks $M$-symmetry. To facilitate the interpretation of the numerical result, we now change to the same notation used in [SM1]: $A_{1}\rightarrow A$, $A_{3}\rightarrow B$. The complex entries $x_{ij}=|x_{ij}|\text{exp}(i\theta_{x_{ij}})$, $x=a,b$, $i,j=1,2$, of these matrices form a set of sixteen real parameters defining the code input. In [SM1] we also define the parameters
%%%%%%%%%%%%%%%%%%%%%%%%%%%%%%%%
\begin{align}
\nonumber p\,=\,&\text{det}A\,\text{det}B\,=\,|p|\,e^{i\alpha},\\
\nonumber q\,=\,&2\sum_{x\neq y}\text{Re}(x_{11}x_{22}^{\ast}y_{12}y_{21}^{\ast})\,-\,2\text{Re}(a_{11}a_{22}^{\ast}b_{11}b_{22}^{\ast})\\
\nonumber    -\,&2\text{Re}(a_{12}a_{21}^{\ast}b_{12}b_{21}^{\ast})\,-\,4\sum_{x\neq y}\text{Im}(x_{11}x_{12}^{\ast})\text{Im}(y_{21}y_{22}^{\ast}),
\end{align}
%%%%%%%%%%%%%%%%%%%%%%%%%%%%%%%%%%%%%%%%%%%%
with $x,y=a,b$. By solving the equation $\text{det}[Q(k)]=0$ in [SM1], we find that the spectrum exhibits a pair of accidental nodes at $k\,=\,k_{\pm}\,=\,\pm\theta+\alpha$, with $\theta\,=\,\arctan(\sqrt{1-t^{2}}/t)$ if $0\leq t\leq1$ or $\theta\,=\,\arctan(\sqrt{1-t^{2}}/t)+\pi$ if $-1\leq t<0$, where $t\equiv-q/(2|p|)$. This is illustrated by the sequence of snapshots obtained from the numerics, as can be seen in \href{http://www.pedebrito.unb.br/index.php?option=com_content&view=article&id=60:symmorphic&catid=38:band&Itemid=67}{Movie 1}. The movie shows how the nodes wander about in the BZ as the parameters $\alpha$ and $t$ are varied. When the condition $t\in[-1,1]$ is not satisfied, the nodes get lifted, i.e. they are accidental.

Applying the same numerical method, we obtain the spectrum of the spinful $STM$-invariant chain described by the Bloch matrix (\ref{HkSM}) with $Q(k)$ given by Eq. (\ref{Q4sSpinfulSM}), and $(-i\sigma_{y})A_{n}(i\sigma_{y})\,=\,A^{\ast}_{n}$, $n=1,3$. As before, we make the notational change $A_{1}\rightarrow A$, $A_{3}\rightarrow B$. The complex entries of $A$ and $B$ are thus constrained by $x_{22}=x_{11}^{\ast}$, $x_{21}=-x_{12}^{\ast}$, $x=a,b$, and hence the code input is defined by eight independent real parameters. In [SM1] we showed analytically that the spectrum of the spinful $STM$-invariant chain displays a symmetric pair of \emph{movable but not removable} degeneracies. This is illustrated in \href{http://www.pedebrito.unb.br/index.php?option=com_content&view=article&id=60:symmorphic&catid=38:band&Itemid=67}{Movie 2} (\href{http://www.pedebrito.unb.br/index.php?option=com_content&view=article&id=60:symmorphic&catid=38:band&Itemid=67}{Movie 3}) which shows how the band structure of the spinful $STM$-invariant chain with $r=4$ changes when varying the phase $\theta_{a_{11}}$ (modulus $|a_{11}|$) of the parameter $a_{11}$, with the other parameters fixed at the same values as in Fig. 4 in [SM1]. We note that the four complex parameters $x_{11}$ and $x_{12}$, $x=a,b$, are on equal footing in the way they influence the location of the nodes, so it is enough to look at the behavior with respect to the phase and modulus of only one of them. In \href{http://www.pedebrito.unb.br/index.php?option=com_content&view=article&id=60:symmorphic&catid=38:band&Itemid=67}{Movie 2}, we see that varying $\theta_{a_{11}}$ in the $[0,2\pi]$ interval makes the nodes to bounce back and forth between the center and the boundaries of the BZ. As shown in \href{http://www.pedebrito.unb.br/index.php?option=com_content&view=article&id=60:symmorphic&catid=38:band&Itemid=67}{Movie 3}, varying $|a_{11}|$ instead causes the nodes to initially approach each other but, past a point of maximum proximity, they are pushed apart and asymptotically move towards their initial positions, as expected (see Fig. 4 in [SM1]. \href{http://www.pedebrito.unb.br/index.php?option=com_content&view=article&id=60:symmorphic&catid=38:band&Itemid=67}{Movie 4} (\href{http://www.pedebrito.unb.br/index.php?option=com_content&view=article&id=60:symmorphic&catid=38:band&Itemid=67}{Movie 5}) shows the same band structure when varying $\theta_{a_{11}}$ ($|a_{11}|$), but with the other parameters fixed at a different set of values. These latter movies are provided in order to illustrate that the crucial feature - namely, that the nodes do not merge at the center or at the boundaries of the BZ - does not rely on a particular choice of the parameters.
\\ \\ \\
$\mbox{[SM1]}$ M. Malard, P. E. de Brito, S. \"Ostlund, and H. Johannesson, Accompanying Letter (2018). \\
$\mbox{[SM2]}$ J. K. Asb\'oth, L. Oroszl\'any and A. P\'alyi, {\em A Short Course on Topological Insulators} (Springer-Verlag, Berlin-Heidelberg, 2016). \\
$\mbox{[SM3]}$ Ching-Kai Chiu, Jeffrey C. Y. Teo, Andreas P. Schnyder and Shinsei Ryu, Rev. Mod. Phys. {\bf 88}, 035005 (2016).

\end{widetext}

\end{document}